\documentclass[a4paper,11pt]{article}
\pdfoutput=1 % if your are submitting a pdflatex (i.e. if you have
\usepackage{jheppub} % for details on the use of the package, please
\usepackage[T1]{fontenc} % if needed
\usepackage{url}

\title{Entanglement Renyi entropy of two disjoint intervals for large $c$ Liouville  field theory }

\author[a]{Jun Tsujimura}
\emailAdd{tsujimura.jun.m5@s.mail.nagoya-u.ac.jp}
\author[a]{Yasusada Nambu}
\emailAdd{nambu@gravity.phys.nagoya-u.ac.jp}

\affiliation[a]{Department of Physics,\\ 
Nagoya University, Chikusa, Nagoya 464-8602, Japan}

\abstract{
     Entanglement entropy (EE) is a quantitative measure of the effective degrees of freedom and the correlation between the sub-systems of a physical system.  Using the replica trick, we can obtain the EE by evaluating the entanglement Renyi entropy (ERE). The ERE is a $q$-analogue of the EE and expressed by the $q$ replicated partition function. In the semi-classical approximation, it is apparently easy to calculate the EE because the classical action represents the partition function by the saddle point approximation and we do not need to perform the path integral for the evaluation of the partition function. In previous studies, it has been assumed that only the minimal-valued saddle point contributes to the EE. In this paper, we propose that all the saddle points contribute equally to the EE by dealing carefully with the semi-classical limit and then the $q \to 1$ limit. For example, we numerically evaluate the ERE of two disjoint intervals for the large $c$ Liouville  field theory  with $q \sim 1$. We exploit the BPZ equation with the four twist operators,  whose solution is given by the Heun function. We determine the  ERE by tuning the behavior of the Heun function  such that it becomes consistent with the geometry of the replica manifold.  We find the same two saddle points as previous studies for $q \sim 1$ in the above system. Then, we provide the ERE for the large but finite $c$ and the $q \sim 1$ in case that all the saddle points contribute equally to the ERE. Based on this work, it shall be of interest to reconsider EE in other semi-classical physical systems with multiple saddle points.
}
\arxivnumber{}
\begin{document}
\maketitle
\flushbottom
\notoc

%------------------------------------------------------------------------------------
\section{Introduction}
\label{sec:1}

     Evaluating the effective degrees of freedom of a physical system is a fundamental  problem in phisics. It is helpful to determine the phases of quantum many-body systems or to study the holographic principle which states that the degrees of freedom of a gravitational system are equal to those of a  system that is one dimension lower  compared to the gravitational system. Entanglement entropy(EE) is a  quantitative measure of the effective degrees of freedom and  the correlation  beween the sub-systems of a physical system ; thus, it has been investigated  from viewpoints of thermodynamics, statistical mechanics, and information theory. Generally, the difficulty in estimating the value of EE depends on the complexity of the structure of  a theory or the form of the sub-systems.  Therefore, despite, the difficulty in ecaluating the EE of general quantum field theory, EEs of two-dimensional conformal field theories are well studied  owing to their  abundant symmetries.  Particularly, the global conformal symmetry determines the EE of a single interval regardless of the  intricacies of the theories. However, when we deal with a two disjoint intervals  sub-system, it is hard to evaluate the EE unless it is a simple  theory such as  the free field \cite{Calabrese:2009ez}.

    The entanglement Renyi entropy(ERE) is a $q$-analogue of the EE. The ERE $S_A(q)$ of the sub-system $A$ is defined as
\begin{align}\label{eq:entanglement Renyi entropy}
    S_A(q) = \frac{1}{1-q} \log \mathrm{tr}_A \left(\rho_A^{\ q} \right),
\end{align}
    where $\rho_A$ is the partial density matrix on $A$. The partial density matrix is normalized as $\mathrm{tr}_A \left(\rho_A \right) = 1$, and then the EE can be defined as $S_A = \lim_{q \to 1}S_A(q)$. The ERE is rewritten as 
\begin{align}
    S_A(q) = \frac{1}{1-q} \left(  \log Z_A(q) - q  \log Z \right),
\end{align}
    where $Z_A(q)$ and $Z$  denote the partition function of the $q$-replicated theory and that of the original theory, respectively \cite{Calabrese:2004eu}. The Liouville  conformal field thoery (CFT) has preferable properties for this formulation, which is studied in the context of the non-critical string theory, higher dimensional theory, etc. \cite{Harlow:2011ny}. The Liouville CFT  exhibits the semi-classical limit as the large $c$ limit. In this limit,  the evaluation of EREs is easier because the saddle points of the path integral represent the respective partition functions.  Previous studies have reported that there exist two saddle points for $Z_A(q)$ for the two disjoint intervals system  in the case of the large $c$ Liouville CFT with $q \sim 1$, or in the adjacent interval limit \cite{Faulkner:2013yia,Hartman:2013mia,Asplund:2014coa}.  Then, it has been assumed that only the minimal valued saddle point contributes to $Z_A(q)$.

    In this paper, we  numerically calculate the ERE for $q \sim 1$  using the Heun function. The Liouville CFT has  postulates that the correlation functions with the null vector  satisfy the linear differential equation known as the BPZ equation prefixed with Belavin, Polyakov and Zamolodchikov \cite{Belavin:1984vu}. As the replica partition function $Z_A(q)$ is given by the correlation function of the twist operators,  this correlation function  can be obtianed by solving the BPZ equation.  Further, we show that the solution  is consistent with the structure of the sub-system. For the two disjoint intervals, the BPZ equation is equivalent to the Heun's differential equation. We determine the ERE by imposing an appropriate condition on the monodromy matrices  of the Heun's differential equation, and  find the two saddle points  that were obtained by the previous studies \cite{Faulkner:2013yia,Hartman:2013mia,Asplund:2014coa}. However, we will point out that  these two saddle points should be treated carefully when  applying the $q \to 1$ limit  for the large $c$, because they contribute equally to $Z_A(q)$,  which can be understood by considering the quantum state corresponding to multiple saddle points.

     This paper is  structured as follows. In section~\ref{sec:2}, we will review the replica trick and  establish the relationship between the geometry of the replica manifold and the correlation function  related to the ERE. In section~\ref{sec:3}, we discuss how to treat multiple saddle points. In section~\ref{sec:4}, we show the EREs for $q \sim 1$ based on the  conditions specified in section~\ref{sec:3}. Finally, section~\ref{sec:5} is the conclusion.

%------------------------------------------------------------------------------------
\section{Entanglement Renyi entropy  (ERE) and replica trick}
\label{sec:2}

\begin{figure}
\begin{center}
    \includegraphics[width=0.9\linewidth]{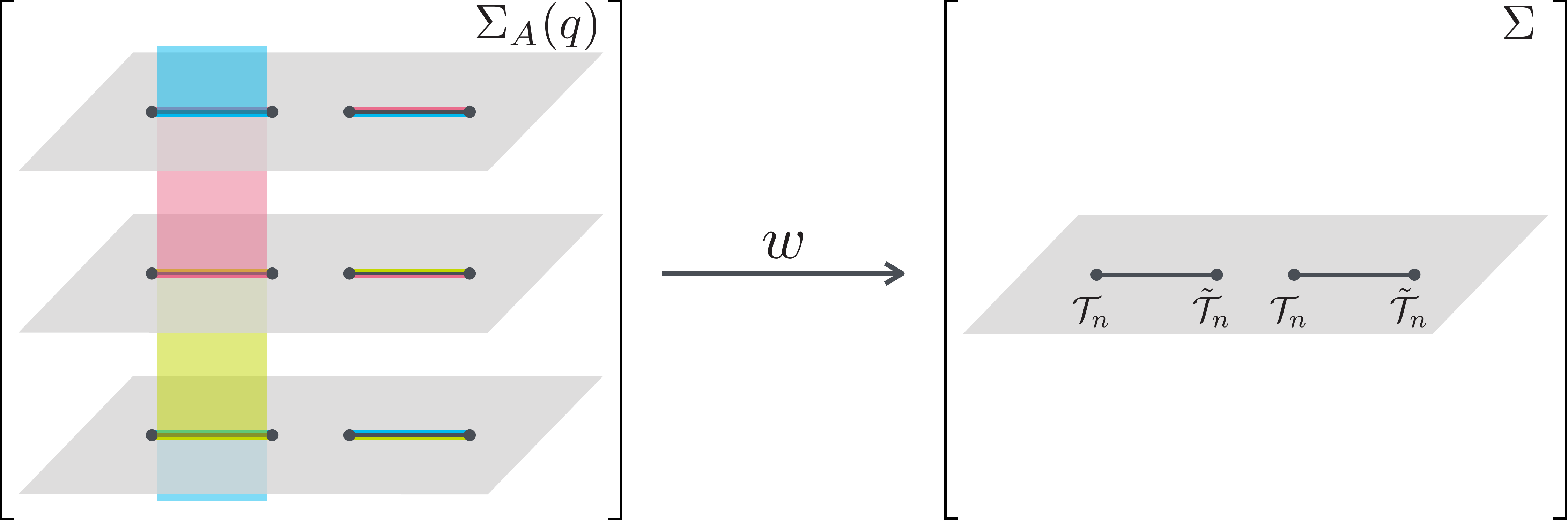}
    \caption{
    The left panel depicts the replica manifold $\Sigma_A(q)$ with $q=3$ sheets. Each sheet is connected to the same -colored lines. The right panel depicts the original manifold $\Sigma$ with the twist operators $\mathcal{T}_q, \mathcal{\tilde{T}}_q$ at the boundary of the sub-system. The conformal map $w$ maps $\Sigma_A(q)$ to $\Sigma$.
    \label{fig:ReplicaManifold}
    }
\end{center}
\end{figure}

    We will review the replica trick and the ERE of two disjoint intervals $A = [z_1,z_2] \cup [z_3,z_4]$ for a $2$-dimensional CFT on the extended complex plane $\Sigma = \mathbb{C} \cup \{ \infty \}$ \cite{Calabrese:2004eu, Calabrese:2009qy}. To evaluate $\mathrm{tr}_A \left(\rho_A^{\ q} \right)$, it is useful to consider the replica manifold and the replica field theory. Fig.~\ref{fig:ReplicaManifold} shows a schematic picture of the replica manifold $\Sigma_A(q)$ of the ERE for two disjoint intervals, the original manifold $\Sigma$ with the twist operators $\mathcal{T}_q, \mathcal{\tilde{T}}_q$, and the conformal map $w: \Sigma_A(q) \to \Sigma$.  The left panel depicts the replica manifold $\Sigma_A(q)$ which  comprises $q$ sheets  and  a single field. The right panel  depicts the replica field theory defined on $\Sigma$, which  comprises $q$ fields on the single sheet with the twist operators.  The replica field theory provides the equivalent partition function to that of the theory on the replica manifold. 
    Let $Z$ be the partition function of the CFT on $\Sigma$, and $Z_A(q)$ be partition function of the same CFT on the replica manifold $\Sigma_A(q)$.  Because $Z_A(q)$ is constituted to satisfy $\mathrm{tr}_A \left(\rho_A^{\ q} \right) =  Z_A(q)/Z^q$, the ERE $S_A(q)$ is  calculated using the partition functions as  follows:
\begin{align}\label{eq:replica trick}
    S_A(q) = \frac{1}{1-q} \left(  \log Z_A(q) - q \log Z \right).
\end{align}
    The replica field theory is constructed so that the above ERE is expressed as the following $4$-point correlation function on $\Sigma$ :
\begin{align}\label{eq:n replicated partition function}
    S_A(q) = \frac{2}{1-q} \log \langle \mathcal{T}_q(z_1)\mathcal{\tilde{T}}_q(z_2)\mathcal{T}_q(z_3)\mathcal{\tilde{T}}_q(z_4) \rangle_{\Sigma},
\end{align}
    where $\mathcal{T}_q$ and $\mathcal{\tilde{T}}_q$  represent the primary twist operators with the same conformal weight $h_q = c(q^2-1)/(24q)$.  Because we deal only with the holomorphic part of the $4$-point correlation function, we multiplied it by the factor $2$ in advance.

    For  a $2$-dimensional CFT, there are  some preferable properties to determine the correlation function. On  the original manifold $\Sigma$, a correlation function  incorporating the energy momentum tensor $T(z)$, and the primary operators $O_i(z_i)$ with the conformal weight $h_i$ satisfies  the following relation:
\begin{align}\label{eq:correlation with emt}
    \langle T(z) \prod_{i=1}^N O_i(z_i) \rangle_{\Sigma} = \sum_{i=1}^N \left[ \frac{h_i}{(z-z_i)^2} + \frac{\partial_{z_i}}{z-z_i} \right] \langle \prod_{i=1}^N O_i(z_i) \rangle_{\Sigma}.
\end{align}
    For an arbitrary operator $\mathcal{O}(z)$, the following relation holds from the definition of the twist operators. 
\begin{align}\label{eq:correlation with twist}
    \frac{\langle \mathcal{O}(z) \mathcal{T}_q(z_1)\mathcal{\tilde{T}}_q(z_2)\mathcal{T}_q(z_3)\mathcal{\tilde{T}}_q(z_4) \rangle_{\Sigma}}{\langle \mathcal{T}_q(z_1)\mathcal{\tilde{T}}_q(z_2)\mathcal{T}_q(z_3)\mathcal{\tilde{T}}_q(z_4) \rangle_{\Sigma}} = q \langle \mathcal{O}(z) \rangle_{\tilde{\Sigma}_A(q)},
\end{align}
    where $\tilde{\Sigma}_A(q)$ is one of the $q$ sheets of the replica manifold $\Sigma_A(q)$ and we use the same complex coordinate on both $\tilde{\Sigma}_A(q)$ and $\Sigma$. If we obtain the conformal transformation $w(z):\ \Sigma_A(q) \to \Sigma$, the energy momentum tensor on the replica manifold is given by the Schwarzian derivative of $w(z)$ as  follows:
\begin{align}\label{eq:Schwarzian derivative}
    \langle T(z) \rangle_{\tilde{\Sigma}_A(q)} = \frac{c}{12}\left[ \frac{w'''(z)}{w''(z)} - \frac{3}{2}\left(\frac{w''(z)}{w'(z)}\right)^2 \right].
\end{align}
    From Eq.~\eqref{eq:correlation with emt}, \eqref{eq:correlation with twist} and \eqref{eq:Schwarzian derivative}, we obtain the following differential equation which relates the conformal transformation, the conformal weight, and the $4$-point correlation function  as follows: 
\begin{align}
    \frac{q c}{12}\left[ \frac{w'''(z)}{w''(z)} - \frac{3}{2}\left(\frac{w''(z)}{w'(z)}\right)^2 \right] = \sum_{i=1}^4 \left( \frac{h_q}{(z-z_i)^2} - \frac{c_i}{z-z_i} \right),
\end{align}
    where $c_i = -\partial_{z_i} \log \langle  \mathcal{T}_q(z_1)\mathcal{\tilde{T}}_q(z_2)\mathcal{T}_q(z_3)\mathcal{\tilde{T}}_q(z_4) \rangle_{\Sigma}$. The global conformal symmetry restricts the correlation function as
\begin{gather}
    \langle \mathcal{T}_q(z_1)\mathcal{\tilde{T}}_q(z_2)\mathcal{T}_q(z_3)\mathcal{\tilde{T}}_q(z_4) \rangle_{\Sigma} = (z_3-z_1)^{-h_q}(z_4-z_2)^{-h_q} \langle  \mathcal{T}_q(0)\mathcal{\tilde{T}}_q(x)\mathcal{T}_q(1)\mathcal{\tilde{T}}_q(\infty) \rangle_{\Sigma} \\
    \sum_{i} c_i = 0,\ \ \ \sum_{i} c_i z_i = 4 h_q,\ \ \ \sum_{i} c_i z_i^2 = 2 h_q \sum_{i} z_i,
\end{gather}
    where $x=(z_4-z_3)(z_2-z_1)(z_4-z_2)^{-1}(z_3-z_1)^{-1}$  denotes one of the cross ratios. Therefore, it is enough to deal with the following equation:
\begin{align}\label{eq:ODE}
    &\frac{w'''(z)}{w''(z)} - \frac{3}{2}\left(\frac{w''(z)}{w'(z)}\right)^2 = \frac{12 h_q}{q c}Q(q,z),\\
    &Q(q,z) = \frac{1}{z^2}+\frac{1}{(z-x)^2}+\frac{1}{(z-1)^2} +2
   \left(\frac{1}{z}-\frac{1}{z-1}\right) + \frac{2 a(q,x)}{z(z-x)(z-1)},
\end{align}
    where $a(q,x)$ is defined as:
\begin{align}\label{eq:derivative of action}
    \frac{ 2 h_q\, a(q,x)}{x(1-x)} = -\partial_x \log \langle  \mathcal{T}_q(0)\mathcal{\tilde{T}}_q(x)\mathcal{T}_q(1)\mathcal{\tilde{T}}_q(\infty) \rangle.
\end{align}
    The solution of the third order non-linear differential equation \eqref{eq:ODE} for $w(z)$ is described as:
\begin{align}\label{eq:conformal map}
    w(z) = \frac{\alpha \Psi_1(z) + \beta \Psi_2(z)}{\gamma \Psi_1(z) + \delta \Psi_2(z)},\ \ \  \alpha \delta - \beta \gamma = 1,
\end{align}
    where $\Psi_1(z),\Psi_2(z)$ are the linearly independent solutions of the following linear differential equation :
\begin{align}\label{eq:Linearized equation}
    \frac{d^2}{dz^2} \Psi(z) + \frac{6 h_q}{q c}Q(q,z) \Psi(z) = 0.
\end{align}
    We can confirm that Eq.~\eqref{eq:conformal map} is the solution of Eq.~\eqref{eq:ODE} by substituting it.  Therefore, the ERE is equivalent to the correlation function of the twist operators, the energy momentum tensor on the replica manifold, the conformal map $w(z):\ \Sigma_A(q) \to \Sigma$,  and the linearly independent solutions of Eq.~\eqref{eq:Linearized equation}.  Thus, the ERE can be obtained by evaluating  any one of these entities. Next, we will evaluate $a(q,x)$ for the semi-classical Liouville CFT. The function $a(q,x)$ is  comparable to the derivative of the ERE. In what follows, we call $a(q,x)$  as the derivative of the ERE. To evaluate the derivative of the ERE $a(q,x)$ for the semi-classical Liouville CFT, we solve Eq.~\eqref{eq:Linearized equation} with the condition that $\Psi(z)$ goes  to the next or previous sheet when crossing the sub-region $A$, as depicted in the left panel of Fig.~\ref{fig:ReplicaManifold}.  Considering the twist operators, this condition  implies that $\Psi(z)$ is a $q$ valued function on $\Sigma$ and the phase of $\Psi(z)$ varies  with $\pm 2\pi/n$ when $\Psi(z)$ goes around the twist operators $\mathcal{T}_q$ and $\mathcal{\tilde{T}}_q$.

    As a practice of the above procedure, let us consider the ERE of the single interval $A = [u,v]$. In this system, two twist operators are inserted at $z_1=u$ and $z_2=v$. From the global conformal symmetry, we can immediately obtain $\langle \mathcal{T}_q(u)\mathcal{\tilde{T}}_q(v) \rangle_{\Sigma} \propto (u-v)^{-2h_q}$ without any other conditions.  Subsequently, we obtain the derivative of the ERE $c_1=-c_2 = 2h_q \, (u-v)^{-1}$ from the definition $c_i = - \partial_{z_i} \log \langle \mathcal{T}_q(u)\mathcal{\tilde{T}}_q(v) \rangle_{\Sigma}$ and the linearized equation corresponding to Eq.~\eqref{eq:Linearized equation} as:
\begin{align}
    &\frac{d^2}{dz^2} \Psi(z) + \frac{6 h_q}{q c}Q(q,z) \Psi(z) = 0,\\
    &Q(q,z) = \frac{1}{(z-u)^2}+\frac{1}{(z-v)^2} - 
   \frac{2(u-v)^{-1}}{z-u} + \frac{2(u-v)^{-1}}{z-v}.
\end{align} 
     The solution for this equation and the corresponding conformal map  are determined as follows: 
\begin{align}
    &\Psi(z) = (z-u)^{ \frac{1}{2}\left(1 \pm \sqrt{1-\frac{24 h_q}{q c}}\right) } (z-v)^{ \frac{1}{2}\left(1 \mp \sqrt{1-\frac{24 h_q}{q c}}\right) },\\
    &w(z) = \frac{\alpha (z-u)^{ \sqrt{1-\frac{24 h_q}{q c}} }+ \beta (z-v)^{ \sqrt{1-\frac{24 h_q}{q c}} }}{\gamma (z-u)^{ \sqrt{1-\frac{24 h_q}{q c}} } + \delta (z-v)^{ \sqrt{1-\frac{24 h_q}{q c}} }}. 
\end{align}
     Because $\Psi(z)$ and $w(z)$ are $q$ valued functions on $\Sigma$, the local behavior of the conformal map should  be consistent with $w(z \sim u) \sim (z-u)^{\pm 1/q}$ ; we further find the conformal weight $h_q = c(q^2-1)/(24q)$ again from the  condition $\pm 1/q = \sqrt{1-24 h_q/(q c)}$. Note that if $\alpha = \delta = 1,\  \beta = \gamma = 0$, we retrieve the well known conformal map $w(z) = (z-u)^{1/q}(z-v)^{-1/q}$. This method works extraordinarily in this example because the behavior of the solutions $\Psi(z)$ is completely determined by the conformal weight of the twist operators  owing to the global conformal symmetry. However,  because general multi point correlation functions depend on  the characteristics of each CFT, we can at most determine the local behavior of $\Psi(z)$ without  applying any other conditions related to the global structure of $\Sigma$. Therefore,  an additional condition  is required to be imposed to determine the global behaviour of $\Psi(z)$. We evaluate the ERE for the large $c$ Liouville CFT on the condition that the $\Psi(z)$ in Eq.~\eqref{eq:Linearized equation}  serves as the $1$-point correlation function on the replica manifold.

%------------------------------------------------------------------------------------
\section{  ERE with multiple saddle points }
\label{sec:3}
    
     In this section, we discuss the treatment of the ERE in the semi-classical approximation , within, the saddle points of the partition function represent the path integral in Eq.~\eqref{eq:replica trick}. According to previous studies \cite{Faulkner:2013yia,Hartman:2013mia,Asplund:2014coa}, the derivative of the ERE is given by $a(q \sim 1,x) = 1-x,\ -x$ from Eq.~\eqref{eq:Linearized equation} for the large $c$ Liouville CFT. We often come across the statement \cite{Dong:2016hjy} that the leading term of the derivative of the ERE for the large $c$ limit is proportional to $(q-1)c$, then $(q-1)c$ must be large enough for the saddle point approximation, and only the minimal -valued action contributes to the path integral for the partition function. However, we  show that all the saddle points may comparably contribute to the ERE for $q \sim 1$. At least, we point out that only one of them does not represent the EE. First, we consider the case  of two saddle points for two disjoint intervals system. From the saddle point approximation, the partition function $Z_A(q)$ is described as  follows:
\begin{align}  
    Z_A(q) = p_1 \exp\left[ -I_{A,1}(q,x) \right] + p_2 \exp\left[ -I_{A,2}(q,x) \right],
\end{align}
    where $p_1,p_2$ are some constants and $I_{A,1}(q,x),\ I_{A,2}(q,x)$ are classical actions. Even after taking the large $c$ limit, the normalization condition $\lim_{q \to 1} \mathrm{tr}_A \left( \rho_A^{\ q} \right) = 1$ must hold. This means $\lim_{q \to 1} (Z_A(q) / Z^q) = 1$, and then
\begin{align}\label{eq:normalization}
     \lim_{q \to 1} \left( \, p_1 \exp\left[ qI -I_{A,1}(q,x) \right] + p_2 \exp\left[ qI -I_{A,2}(q,x) \right] \,\right)= 1,
\end{align}
    where we assume that $I = - \log Z$ is the unique Euclidean classical action of the original theory ; it is  the $c$  order term.  Because the replica field theory  involves the $q$ replicated field of the original theory, the effective action $I_{A,i}(q,x)$ may  be comparable to $I_{A,i}(q,x) = q I + O(q-1)$ for $q \sim 1$. Thus, $p_1 + p_2 = 1$ from Eq.~\eqref{eq:normalization}. One may concern that the two saddle points merge into the  saddle point of the original theory for $q \sim 1$ and become indistinguishable. Although the values of the two classical actions are infinitely close in  the $q \to 1$ limit, the two saddle points are distinct  considering the path integral because the configuration of the classical fields $\Psi(z)$ are topologically different as  observed in previous studies. Thus, there exist the two distinct saddle points,  out of which   the leading term behaves like  the $c$  order term, as long as $q \neq 1$. Then, each classical action are described as $I_{A,i}(q,x) = q I + b_i(q,x)$, $\lim_{q \to 1} b_i(q,x) = 0$. Therefore, Eq.\eqref{eq:replica trick} for an arbitrary $q$ in the semi-classical limit becomes 
\begin{align}\label{eq:ERE with multi saddles}
    S_A(q) = \frac{1}{1-q} \left(  \log \sum_{i=1}^2 p_i \exp \left[-b_i(q,x) \right] \right).
\end{align}
    As a result, the term in the parenthesis is proportional to $(q-1)c$. However, note that the leading terms of the saddle points for $Z_A(q)$ are proportional to $q I$, and then Eq.\eqref{eq:ERE with multi saddles} is derived from the cancellation between $q I$,  which originated from $\log Z_A(q)$ and one from $q \log Z$. Thus, the saddle point approximation is valid for  a large $c$ independent of  the magnitude of $(q-1)c$.

    We obtained Eq.\eqref{eq:ERE with multi saddles} as the ERE in the semi-classical limit with an arbitrary $q$  based on the assumption that the replica field theory has two saddle points for the large $c$. If we adopt  a large enough but finite $c$ for the saddle point approximation and keep $q-1$ finite, we can  consider that only the minimal saddle point contributes to the ERE for  a large $(q-1)c$.  Conversely, if $(q-1)c \sim 0$ with large finite $c$ and $q \sim 1$, the ERE becomes
\begin{align}
    S_A(q \sim 1) \sim \frac{1}{1-q} \log  \sum_{i=1}^2 p_i \left( 1-b_i (q,x) \right) 
    \sim \frac{1}{1-q} \left[\log \sum_{i=1}^2 p_i - \sum_{i=1}^2 \frac{p_i b_i(q,x)}{ \sum_{i=1}^2 p_i }\right].
\end{align}
     Owing to the normalization condition  of $p_1+p_2=1$, the EE is  defined as:
\begin{align}\label{eq:entanglement entropy}
    S_A = \lim_{q \to 1} S_A(q) = \lim_{q \to 1} \sum_{i=1}^2 \frac{p_i\, b_i(q,x)}{q-1}.
\end{align}
     Thus, the EE  determined in the semi-classical limit is  a summation of all  the $(q-1)c$  order terms of the classical actions.  The following two nuances should be noted: First, in the semi-classical approximation, the leading terms of the classical action of the two partition functions cancel each other  owing to the structure of the replica theory and the normalization condition of the density matrix.  Second, the $q \to 1$ limit is  adopted so that $(q-1) c \sim 0$ is satisfied. Therefore, because of the exquisite relationship between the two limits, the multiple saddle points equally contribute to the EE. The above cautions are specific to the EE in the semi-classical approximation. Thus,  we do not need to concern about it in other  scenarios, such as the thermal phase transition of physical systems.

    We identify the relation between the derivative of the ERE $a(q,x)$ and the order $(q-1)c$ term of the classical action $b_i(q,x)$, and then determine $p_1$ and $p_2$.  We should pay attention for the quantum state of the replica field theory to relate them. As there are two classical saddle points, it is natural that the replica field theory also has the two quantum states corresponding to respective them. We assume that the quantum state of the replica field theory is expressed as $|\Omega\rangle = \sqrt{p_1} |\Omega_1\rangle  + \sqrt{p_2} |\Omega_2\rangle$, $\langle \Omega_1|\Omega_2\rangle \sim 0$, where $|\Omega_1\rangle$ and $|\Omega_2\rangle$  represent the states corresponding to the respective classical actions in the semi-classical limit.  Subsequently, the partition function is described as $Z_A(q) = p_1 Z_{A,1}(q) + p_2 Z_{A,2}(q)$, where we defined $Z_A(q) = \langle \Omega|\Omega\rangle,\ Z_{A,i}(q) = \langle \Omega_i|\Omega_i\rangle$. We can  associate the weights $p_1$ and $p_2$ to the probability amplitude of $|\Omega_1\rangle$ and $|\Omega_2\rangle$, respectively. According to the above argument, Eq.~\eqref{eq:derivative of action} is written as:
\begin{align}
    \frac{ 2 h_q }{x(1-x)} a(q,x) = -\frac{p_1 \partial_x Z_{A,1}(q) + p_2 \partial_x Z_{A,2}(q)}{Z_A(q)} = -\frac{p_1 \partial_x e^{-b_1(q,x)} + p_2 \partial_x e^{-b_2(q,x)}}{p_1 e^{-b_1(q,x)}+ p_2 e^{-b_2(q,x)}}.
\end{align}
     Because the ERE for $q \sim 1$ is  equivalent tothe summation of the derivative of the classical actions, as  described in Eq.~\eqref{eq:entanglement entropy}, it  is natural that the derivative of the ERE also  decomposes into $a(q,x) = p_1 a_1(q,x) + p_2 a_2(q,x)$ at least for $q \sim 1$. In particular, we relate $a_i(q,x)$ and $b_i(q,x)$ as  follows:
\begin{align}\label{eq:a and b}
    \frac{ 2 h_q }{x(1-x)} a_i(q,x) = - \frac{\partial_x e^{-b_i(q,x)}}{p_1 e^{-b_1(q,x)}+ p_2 e^{-b_2(q,x)}} \sim \partial_x b_i(q\sim 1,x)
\end{align}
     Note that the two candidates of the derivative of the ERE $a(q \sim 1,x) = 1-x,\ -x$ are obtained just by analyzing Eq.~\eqref{eq:Linearized equation} independent of the quantum state. Therefore, we assume
\begin{align}
    p_1 a_1(q \sim 1,x) = 1-x, \ \ \  p_2 a_2(q \sim 1,x) = -x.
\end{align}
    Next, we determine the probability amplitude $p_1$ and $p_2$ because we can only obtain $p_i a_i(q,x)$  and not $a_i(q,x)$ itself.  Furthermore, the ERE is described by the $4$-point function of the twist operators Eq.~\eqref{eq:n replicated partition function}. Consider the $4$-point correlation function $G_{1234}(x) = \langle \phi_1(0)\phi_2(x)\phi_3(1)\phi_4(\infty) \rangle_{\Sigma}$, where $\phi_i$ is a general operator.  Because $G_{1234}(x)$ is independent of the way of the operator product expansion,  and $G_{1234}(x)$  exhibits the crossing symmetry $G_{1234}(x) = G_{3214}(1-x)$.  The first and the third operators, in the ERE in Eq.~\eqref{eq:n replicated partition function},  are identical twist operators; therefore, we obtain $G_{1234}(x) = G_{3214}(x)$  in addition to $G_{1234}(x) = G_{1234}(1-x)$. Therefore, the ERE $S_A(q)$ in Eq.~\eqref{eq:ERE with multi saddles} is invariant with the replacement $x \to 1-x$; thereby, allowing $p_1 = p_2 = 1/2$  to true. Finally, the EE of the two disjoint intervals for the large $c$ Liouville CFT from Eq.~\eqref{eq:entanglement entropy} is
\begin{align}  
    S_A = \lim_{q \to 1} \frac{4 h_q}{q-1} \int \frac{ 1-2x}{x(1-x)} dx  = \frac{c}{3} \log \frac{x(1-x)}{\epsilon^2},
\end{align}
    where $\epsilon$  denotes the UV cut off scale.  Consequently, the obtained EE is equivalent to that of the free compactified boson at the leading order of the large $c$. Thus, it  shall not be in contradiction to any  postulate of  the CFT. Note that we do not need the weights $p_i$ to calculate the EE, and we exploited the symmetry between the two saddle points to determine the weights $p_i$. However, if there are more than two saddle points, we need some extra information to evaluate the weights $p_i$ and the ERE.

%------------------------------------------------------------------------------------
\section{  Determination of ERE for the semi-classical Liouville CFT}
\label{sec:4}

     In this section, we see that $\Psi(z)$ in Eq.~\eqref{eq:Linearized equation} should behave as the $1$-point correlation function on the replica manifold for the large $c$ Liouville CFT,  and then determine the ERE of the two disjoint intervals. For the Liouville CFT, the BPZ equation helps us to analyze the structure of the correlation functions. Let $\psi_\chi(z)$ denote the operator corresponding  to the level $2$ light null vector with the conformal weight $h_{\chi}$ ; wherein, the following BPZ equation holds :
\begin{align}\label{eq:BPZ equation}
    \left[ \frac{3 \partial^2_z}{2(2h_{\chi}+1)} - \sum_{i=1}^4 \left( \frac{h_q}{(z-z_i)^2} + \frac{\partial_{z_i}}{z-z_i} \right) \right] \langle \psi_\chi(z) \mathcal{T}_q(z_1)\mathcal{\tilde{T}}_q(z_2)\mathcal{T}_q(z_3)\mathcal{\tilde{T}}_q(z_4) \rangle = 0
\end{align}
    As we treat  the $q$-replicated Liouville CFT, the central charge is $q$ times  that of the original theory, that is, $h_{\chi} = (5-qc+\sqrt{(qc-1)(qc-25)})/16$.  We can choose $(z_1,z_2,z_3,z_4)=(0,x,1,\infty)$ without  the loss of generality. In the large $c$ semi-classical limit, we can rewrite this equation  in a simple form  through the following steps. The conformal weight $h_{\chi}$ is $h_{\chi} = -1/2 - 9/(2qc) + O(c^{-2})$ and $\psi_\chi(z)$ is a light operator  whose expectation value  can be considered as a $1$-point correlation function $\Psi_\chi(z)$ on the replica manifold. This means that the above $5$-point correlation function behaves as  follows:
\begin{align}    
    &\Psi_\chi(z) = \frac{\langle \psi_\chi(z) \mathcal{T}_q(0)\mathcal{\tilde{T}}_q(x)\mathcal{T}_q(1)\mathcal{\tilde{T}}_q(\infty) \rangle_{\Sigma}}{\langle \mathcal{T}_q(0)\mathcal{\tilde{T}}_q(x)\mathcal{T}_q(1)\mathcal{\tilde{T}}_q(\infty) \rangle_{\Sigma}} \\
    &\Longrightarrow \langle \psi_\chi(z) \mathcal{T}_q(0)\mathcal{\tilde{T}}_q(x)\mathcal{T}_q(1)\mathcal{\tilde{T}}_q(\infty) \rangle_{\Sigma} = \Psi_\chi(z) \, \sum_{i} p_i \, e^{-I_{A,i}(q,x)}.
\end{align}
    Thus, we will deal with the following equation assuming that $\Psi_\chi(z)$ behaves as the $1$-point correlation function on the replica manifold: 
\begin{align}\label{eq:large c BPZ}
    &\frac{d^2}{dz^2} \Psi_\chi(z)  + \frac{q^2-1}{4q^2}Q(q,z) \Psi_\chi(z)  = 0,\\
    \label{eq: Q}
    &Q(q,z) = \frac{1}{z^2}+\frac{1}{(z-x)^2}+\frac{1}{(z-1)^2} +2
   \left(\frac{1}{z}-\frac{1}{z-1}\right) + \frac{2 a(q,x)}{z(z-x)(z-1)},\\
   \label{eq: ai}
    &a(q,x) = -\frac{12 q}{c(q^2-1)}x(1-x) \partial_x \log Z(q,x).
\end{align}
    $\Psi_\chi(z)$ satisfies the same differential equation as Eq.~\eqref{eq:Linearized equation}, but now we have  an additional global condition that $\Psi_\chi(z)$ behaves as a $1$-point correlation function on the replica manifold.  Furthermore, we evaluate $a(q,x)$ for $q \sim 0$ first as we can find  an analytical expression of $\Psi_\chi(z)$  using the WKB  approximation, and then numerically evaluate $a(q,x)$ for $q \sim 1$.

    First, as just a practice, we calculate the ERE for $q \sim 0$  using the WKB method because it  enables in for understanding the relation between the structure of the replica manifold and the global behavior of $\Psi_\chi(z)$ on it. Consider the following WKB solution of Eq.~\eqref{eq:large c BPZ} in the leading order of the WKB approximation for $q \sim 0$:
\begin{align}
    \Psi_\chi(z) = \frac{1}{Q(q,z)^{1/4}} \exp \left[ \pm \frac{1}{2q} \int^z \sqrt{Q(q, \zeta)} d\zeta \right].
\end{align}
    As we have the integral expression for $\Psi(z)$, it is easy to analyze its global behavior which is determined by the residues of $\sqrt{Q(q,z)}$. Note that we can rewrite $\sqrt{Q(q,z)}$ as
\begin{align}
    \sqrt{Q(q,z)} = \frac{\sqrt{z^4 -2z^3 +2z^2 -2xz +x^2 + 2 a(q,x) z(z-x)(z-1) }}{z(z-x)(z-1)}.
\end{align}
    The residues of $\sqrt{Q(q,z)}$ at $z=0,x,1$ are $\pm 1$ independent of $a(q,x)$. From the requirement that $\Psi(z)$ behaves as a $1$-point correlation function on the replica manifold as depicted in Fig.\ref{fig:ReplicaManifold}, $a(q,x)$ is determined so that $\sqrt{Q(q,z)}$  transforms into a rational function and its Riemann surface is single sheeted, that is,  $\mathrm{Res}\sqrt{Q(q,z=0)}=-\mathrm{Res}\sqrt{Q(q,z=x)}=\mathrm{Res}\sqrt{Q(q,z=1)}$ should holds. Therefore, we find the unique derivative of the ERE $a(q,x) = 1-2x$, and then, $\sqrt{Q(q,z)}$ and the ERE for $q \sim 0$ is  determined as follows:
\begin{gather}
    \sqrt{Q(q \sim 0, z)} = \pm \left( \frac{1}{z} - \frac{1}{z-x} + \frac{1}{z-1} \right), \\
    S_A(q \sim 0) = \lim_{q \to 0}\frac{4 h_q}{q-1} \int \frac{1-2x}{x(1-x)} dx  = \frac{c}{6q} \log \frac{x(1-x)}{\epsilon^2}.
\end{gather}
    We can express the conformal map as $w(z) = z^{\frac{1}{q}}(z-x)^{-\frac{1}{q}} (z-1)^{\frac{1}{q}}$ ; we obtain the energy momentum tensor for $q \sim 0$ as  follows:
\begin{align}
    \frac{12}{q\,c} T(z)
    &=\frac{q^2-1}{2 q^2 z^2}
    +\frac{q^2-1}{2 q^2 (z-x)^2}
    +\frac{q^2-1}{2 q^2 (z-1)^2} - \frac{\left(q^2-1\right) (x+z-1)}{q^2z (z-x) (z-1) } -\frac{6 (x-1) x}{(z^2 - 2xz + x)^2}\nonumber\\
    &\sim -\frac{1}{2q^2} \left[ \frac{1}{z^2} + \frac{1}{(z-x)^2} + \frac{1}{(z-1)^2} + 2\left(\frac{1}{z} -\frac{1}{z-1}\right) + \frac{2(1-2x)}{z (z-x) (z-1)} \right]
\end{align}
     The form of this energy momentum tensor  is consistent  with Eq.\eqref{eq:correlation with emt} for $q \sim 0$,  that is, it has the same poles. For  a finite $q$, the sub-leading terms of the WKB solution may cancel the extra poles at $z^2 - 2xz + x=0$.  Additionally, for $a(q,x) = \pm 1$, $\sqrt{Q(q,z)}$ also becomes a rational function :
\begin{align}
    &a(q,x) = 1 \ \Longleftrightarrow\ \sqrt{Q(q \sim 0, z)} = \pm \left( \frac{1}{z} - \frac{1}{z-x} - \frac{1}{z-1}\right),\\
    &a(q,x) = -1 \ \Longleftrightarrow\ \sqrt{Q(q \sim 0, z)} = \pm \left( \frac{1}{z} + \frac{1}{z-x} - \frac{1}{z-1}\right).
\end{align}
    From the relative sign of the poles, the $4$-point correlation functions corresponding to them are  given as:
\begin{align}
    &a(q,x) = 1\ \Longleftrightarrow\ \langle \mathcal{T}_q(z_1)\mathcal{\tilde{T}}_q(z_2)\mathcal{\tilde{T}}_q(z_3)\mathcal{T}_q(z_4) \rangle_{\Sigma}, \\
    &a(q,x) = -1 \ \Longleftrightarrow\ \langle \mathcal{T}_q(z_1)\mathcal{T}_q(z_2)\tilde{\mathcal{T}}_q(z_3)\tilde{\mathcal{T}}_q(z_4) \rangle_{\Sigma}.
\end{align}
    This practice clearly demonstrates the relation between each $4$-point correlation function and the geometry of each replica manifold. We may be able to precisely analyze by considering the higher order term of the WKB solution.  Thus, we confirm the one-to-one correspondence between each saddle point $a_i(q,x)$ and each replica manifold. However, we obtain multiple saddle points for  the general $q$.

    Second, we  consider the $q \sim 1$ case. Let $\Phi(z) = g(x)\Psi_\chi(z) $ and $g(z) = z^{\frac{q-1}{2q}}(z-x)^{\frac{q+1}{2q}}(z-1)^{\frac{q-1}{2q}}$, then Eq.~\eqref{eq:large c BPZ} is transformed into the Heun's differential equation  as follows: 
\begin{align}\label{eq:Heun standard}
    &\frac{d^2}{d z^2} \Phi(z) + \left(\frac{\gamma}{z}+\frac{\epsilon}{z-x}+\frac{\delta}{z-1}\right) \frac{d}{d z} \Phi(z) + \frac{\alpha \beta z - p }{z(z-x)(z-1)} \Phi(z) = 0,\\
    &\alpha = 1,\ 
    \beta = 1-\frac{1}{q},\ 
    \gamma = 1-\frac{1}{q},\ 
    \delta = 1-\frac{1}{q},\ 
    \epsilon = 1+\frac{1}{q},\\
    &p = \frac{q-1}{2 q^2} \left[ 1-2x+q-(q+1) a_i(q,x) \right]
\end{align}
    The solution of this equation is called as the Heun function. The Heun's differential equation has four regular singular points at $z=0,x,1,\infty$ and the Frobenius solutions of the Heun's differential equation  are known as the local Heun functions. For example, two independent local Heun functions around $z=0$ can be expressed as :  
\begin{align}  
    &\Phi(z \sim 0) \sim \text{HeunG}[x,p,\alpha ,\beta ,\gamma ,\delta ,z],\\
    &\Phi(z \sim 0) \sim z^{1-\gamma } \text{HeunG}[x,p+(1-\gamma ) (\delta  x+\epsilon ),\alpha -\gamma +1,\beta -\gamma +1,2-\gamma ,\delta ,z],
\end{align}
    where the local Heun function is normalized as $\text{HeunG}[x,p,\alpha ,\beta ,\gamma ,\delta ,z=0]=1$ \cite{HeunG}. We denote a local Heun function near $z=z_i$ with the characteristic exponent $s$ as $y_{z_i}^{s}(z)$. The connection matrix describes  the relationships between the local Heun functions. For example, $y_{x}^{s}(z)$ and $y_{0}^{s}(z)$ are connected by the connection matrix $C_{x0}$ as  follows:
\begin{align}\label{eq:connection matrix}
    \left(  
    \begin{array}{c}
    y_{x}^{0}(z)\\
    y_{x}^{1-\epsilon}(z)
    \end{array}
    \right)
    =
    \frac{1}{W(y_{0}^{0},y_{0}^{1-\gamma})}
    \left(  
    \begin{array}{cc}
    W(y_{x}^{0},y_{0}^{1-\gamma}) & W(y_{0}^{0},y_{x}^{0}) \\
    W(y_{x}^{1-\epsilon},y_{0}^{1-\gamma}) & W(y_{0}^{0},y_{x}^{1-\epsilon})
    \end{array}
    \right)
    \left(  
    \begin{array}{c}
    y_{0}^{0}(z)\\
    y_{0}^{1-\gamma}(z)
    \end{array}
    \right),
\end{align}
    where $W(y_{x}^{0},y_{x}^{1-\epsilon}) = y_{x}^{0}(z)\partial_z y_{x}^{1-\epsilon}(z) - \partial_z y_{x}^{0}(z) y_{x}^{1-\epsilon}(z)$ is the Wronskian of $y_{x}^{0}(z)$ and $y_{x}^{1-\epsilon}(z)$ and the others are the same. The ratio of these Wronskians  attains a constant value with respect to $z$,  contrary to the Wronskians themselves. We utilized the Mathematica to calculate these Wronskians, see \cite{Hatsuda:2020sbn}.

\begin{figure}[ht]
\begin{center}
    \includegraphics[width=0.85\linewidth]{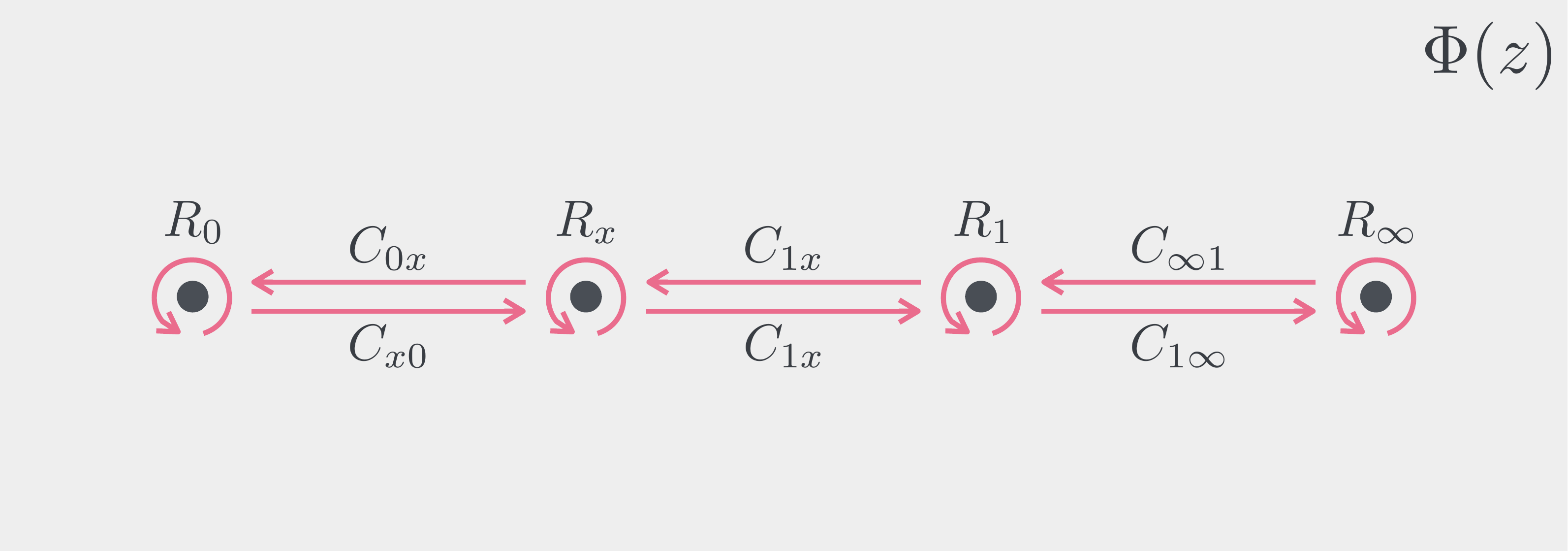}
    \caption{
        The dots represent $z=0,x,1,\infty$ on $\Sigma$ from left to right. The analytic continuation along each magenta line  is described as  a matrix,  such as $C_{x0},\cdots$ and $R_{0},\cdots$. The paths $P_{0x}$ and $P_{1x}$ correspond to the monodromy matrices $M_{0x} = C_{x0}R_{0}C_{0x}R_{x}$ and $M_{1x} = C_{x1}R_{1}C_{1x}R_{x}$, respectively.
        \label{fig:path}
    }
\end{center}
\end{figure}

    The derivative of the ERE $a(q,x)$ determines the connection matrices.  Additionally, we  need to find the condition that the connection matrices must satisfy. To formulate it, consider the paths $P_{0x}$ and $P_{1x}$ which encircle the interval $[0,x]$ or $[x,1]$  once in  the counterclockwise  direction. And let $R_{0} = R_{x}^{-1} = R_{1} = R_{\infty}^{-1} = \mathrm{diag}(1,\exp[2\pi i/q])$ and the connection matrix $C_{x0}$ be given by Eq.~\eqref{eq:connection matrix} and the others be  defined in the same manner. Then, the analytic continuation along $P_{0x}$ for the local Heun functions $y_{x}^{0}$ and $y_{x}^{1-\epsilon}$ are described as :
\begin{align}
    \left(  
    \begin{array}{c}
    y_{x}^{0}(z)\\
    y_{x}^{1-\epsilon}(z)
    \end{array}
    \right)
    =
    M_{0x}
    \left(  
    \begin{array}{c}
    y_{x}^{0}(z)\\
    y_{x}^{1-\epsilon}(z)
    \end{array}
    \right),
\end{align}
    where we define the monodromy matrix $M_{0x} = C_{x0}R_{0}C_{0x}R_{x}$ as depicted in Fig.~\ref{fig:path}. 
     Similarly, the analytic continuation along $P_{1x}$ is expressed  comparable to the other monodromy matrix $M_{1x} = C_{x1}R_{1}C_{1x}R_{x}$.  One may hope that both the monodromy matrices  transforms into the identity matrix like the WKB analysis for $q \sim 0$. However, both cannot transforms into the identity matrix simultaneously for general $q$ whatever $a(q,x)$ is chosen. Instead, one of  two should be the identity matrix,  also known as the Schottky uniformization \cite{Faulkner:2013yia,Krasnov:2000zq,Zograf_1988}.  Moreover, it is trivial  for the analytic continuation along the path which encircles all the four regular singular points  once in  the counterclockwise  direction. Therefore, $M_{0x}=I$ is equivalent to $M_{\infty 1}=I$ because $C_{x1} M_{\infty 1} C_{1x} M_{0x}=I$ and $C_{1x} C_{x1} = I$.  Comparably, $M_{1x}=I$  implies $M_{0 \infty}=I$ ; thus, it is  sufficient to deal with the monodromy matrices $M_{0x}$ and $M_{1x}$. On this condition, the monodromy matrices $M_{0x}$ and $M_{1x}$ are commutative. Thus, if we perform the analytical continuation via $q$ times $P_{0x}$ and $q$ times $P_{1x}$ for integer $q$, $\Phi(z)$  retains its original value  because this is the first time that $\Phi(z)$ is back to the starting point from the viewpoint of the replica manifold. For $0 < q \in \mathbb{Q}$, let $q = t/u$ with $t,u \in \mathbb{Z}^+$,  while considering the analytical continuation via $u$ times $P_{0x}$ and $u$ times $P_{1x}$ in random order, the same discussion holds because $\Phi(z)$ is $t$ times back to the starting point. Therefore, we accept the Schottky uniformization for an arbitrary $q \in \mathbb{R}$.

    For  an arbitrary $q$ near $x=0$ or $x=1$, the derivative of the ERE behaves as $a(q,x \sim 0) \sim 1$ or $a(q,x \sim 1) \sim -1$, respectively \cite{Hartman:2013mia,Asplund:2014coa}. Then, we regard the former as $p_1 a_1(q,x)$ if there are only two saddle points. We numerically calculate $p_1 a_1(q,x)$  in case all the components of the commutation relation between the two monodromy matrices $[ M_{0x},M_{1x} ]$ vanish  simultaneously. Then, we obtain the ERE from Eq.\eqref{eq:ERE with multi saddles} and \eqref{eq:a and b} with $a_1(q,x) = -a_2(q,1-x)$.
\begin{figure}[ht]
\begin{center}
    \includegraphics[width=\linewidth]{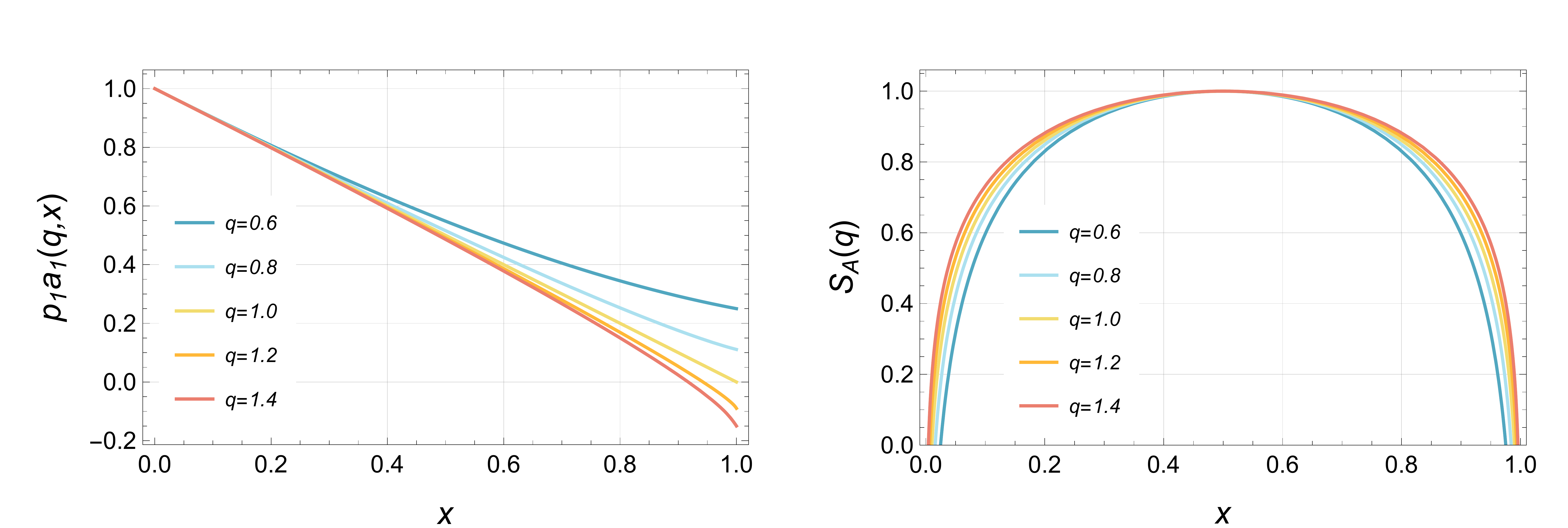}
    \caption{
        The left panel shows $p_1 a_1(q,x)$ for  $q=0.6,0.8, 1.0, 1.2, 1.4$. The right panel shows the corresponding ERE normalized as $S_A(q)=1$ at $x=1/2$ with the central charge $c=1$ and the UV cutoff  $\epsilon=0.1$.
        \label{fig:ERE}
    }
\end{center}
\end{figure}
    Fig.~\ref{fig:ERE} shows $p_1 a_1(q,x)$ and the ERE  $S_A(q)$ for $q \sim 1$. For $q \to 1$, we can  consider $p_1 a_1(q \to 1,x) \to 1-x$. As mentioned before, we obtained the same EE as  that of a compactified boson. Note that the central charge $c$ should be large enough  because Eq.\eqref{eq:ERE with multi saddles} is based on the saddle point approximation. For $(q-1)c \sim 0$, the EREs depend on $c$ only linearly, and then the EREs with $c=1$ in Fig.~\ref{fig:ERE} is meaningful. It is difficult to compute the ERE not for $q \sim 1$. In particular for $q < 0.5$, the number of saddle points  increases with a decreasing $q$, and we cannot determine the weights of the contribution for each saddle point to the ERE. Moreover, we cannot calculate each saddle point for small $q$  owing to the lack of numerical accuracy. The WKB analysis  could be considered to calculate the ERE for this region.

%------------------------------------------------------------------------------------
\section{Conclusion}
\label{sec:5}

    In this study, we reviewed the relationship between the ERE and the geometrical structure of the replica manifold and saw that some additional conditions must be imposed to determine the ERE of two disjoint intervals system in general. Then, we  considered the treatment of the EE in the semi-classical approximation in general. Because of the exquisite relationship between the large $c$ and $q \to 1$, we pointed out that the multiple saddle points contribute equally to the EE. The leading terms of the classical action of the two partition functions $Z_A(q)$ and $Z$ for large $c$ cancel each other due to the structure of the replica theory and the normalization condition of the density matrix. For the  case of general ERE,  the method to evaluate the contribution weights of each saddle point  is not known; and for the EE we do not need to know the weights as shown in Eq.~\eqref{eq:entanglement entropy}.  Thus, we numerically evaluated the ERE of the two disjoint intervals for the large $c$ Liouville CFT for $q \sim 1$ by analyzing the BPZ equation  by satisfying the criterion that its solution behaves like a $1$-point correlation function on a replica manifold. This condition is expressed by the condition that one of the monodromy matrices  transforms into the identity matrix for any real number $q$.

    In future work, it  shall be of interest to reconsider ERE in other  scenarios and entanglement measures. For  instance, there is a growing interest in the reflected Renyi entropy,  which signifies that the corresponding replica manifold  exhibits a rather complex geometry \cite{Dutta:2019gen}.  Additionally, we can consider the ERE of a single intercal on the torus as it is also well known for the expression of the Heun's differential equation. Conversely, it would be interesting to evaluate the higher order terms of the WKB method and the large $c$. Considering the higher order terms of the WKB method for finite $q$, we may check the consistency between the WKB method and the numerical method for  the ERE of the two disjoint intervals. For higher order corrections of large $c$, it may be necessary to evaluate the contribution of the cross term between multiple quantum states corresponding with each saddle point  in case of multiple saddle points.

\begin{acknowledgments}
    The author J.T. was financially supported by JSPS Fellows (22J14390). The author Y.N. was supported in part by JSPS KAKENHI Grant No. 19K03866. 
\end{acknowledgments}

%------------------------------------------------------------------------------------
\bibliographystyle{JHEP}
\bibliography{ERE_of_two_disjoint_intervals_for_large_c_Liouville_CFT}
 
\end{document}